%% file: main.tex
\documentclass[
    a4paper,
    amsmath,amssymb
]{jpconf}

\usepackage{graphicx}
\usepackage{amsmath}
\usepackage{hyperref}
\usepackage[sort&compress,square,comma,numbers]{natbib}
\usepackage{todonotes}

\begin{document}

\title{Reconstruction of Large Radius Tracks with the Exa.TrkX pipeline}

\author{
Chun-Yi Wang\textsuperscript{1},
Xiangyang Ju\textsuperscript{2},
Shih-Chieh Hsu\textsuperscript{3},
Daniel Murnane\textsuperscript{2},
Paolo Calafiura\textsuperscript{2},
Steven Farrell\textsuperscript{2}, 
Maria Spiropulu\textsuperscript{4},
Jean-Roch Vlimant\textsuperscript{4},
Adam Aurisano\textsuperscript{5},
V Hewes\textsuperscript{6}, Giuseppe Cerati\textsuperscript{6}, Lindsey Gray\textsuperscript{6}, Thomas Klijnsma\textsuperscript{6}, Jim Kowalkowski\textsuperscript{6}, 
Markus Atkinson\textsuperscript{7}, Mark Neubauer\textsuperscript{7},
Gage DeZoort\textsuperscript{8}, Savannah Thais\textsuperscript{8},
Alexandra Ballow\textsuperscript{9}, Alina Lazar\textsuperscript{9},
Sylvain Caillou\textsuperscript{10}, Charline Rougier\textsuperscript{10}, Jan Stark\textsuperscript{10}, Alexis Vallier\textsuperscript{10}, Jad Sardain\textsuperscript{10}
}

\address{
\textsuperscript{1}~National Tsing Hua University,
\textsuperscript{2}~Lawrence Berkeley National Lab,
\textsuperscript{3}~University of Washington,
\textsuperscript{4}~California Institute of Technology,
\textsuperscript{5}~University of Cincinnati,
\textsuperscript{6}~Fermi National Accelerator Laboratory,
\textsuperscript{7}~University of Illinois Urbana-Champaign,
\textsuperscript{8}~Princeton University,
\textsuperscript{9}~Youngstown State University,
\textsuperscript{10}~Laboratoire des 2 Infinis - Toulouse (L2IT-IN2P3)
}

\ead{xju@lbl.gov}

\input{chapters/abstract}
\input{chapters/introduction}

\input{chapters/pipeline}
\input{chapters/data_generation}
\input{chapters/result}
\input{chapters/conclusion}

\ack
This research was supported in part by the U.S. Department of Energy’s Office of Science, Office of High Energy Physics, of the US Department of Energy under Contracts No. DE-AC02-05CH11231 (CompHEP Exa.TrkX) and No. DE-AC02-07CH11359 (FNAL LDRD 2019.017); and by the National Science Foundation under Cooperative Agreement OAC-1836650.

This research used resources of the National Energy Research Scientific Computing Center (NERSC), a U.S. Department of Energy Office of Science User Facility located at Lawrence Berkeley National Laboratory, operated under Contract No. DE-AC02-05CH11231.

\bibliographystyle{iopart-num}
\bibliography{main}


\end{document}

%% file: chapters/abstract.tex
\begin{abstract}

Particle tracking is a challenging pattern recognition task at the Large Hadron Collider (LHC) and the High Luminosity-LHC. Conventional algorithms, such as those based on the Kalman Filter, achieve  excellent performance in reconstructing the prompt tracks from the collision points.
However, they require dedicated configuration and additional computing time to efficiently reconstruct the large radius tracks created away from the collision points. 
We developed an end-to-end machine learning-based track finding algorithm for the HL-LHC, the Exa.TrkX pipeline. The pipeline is designed so as to be agnostic about global track positions. In this work, we study the performance of the Exa.TrkX pipeline for finding large radius tracks.
Trained with all tracks in the event, the pipeline simultaneously reconstructs prompt tracks and large radius tracks with high efficiencies.
This new capability offered by the Exa.TrkX pipeline may enable us to search for new physics in real time.

\end{abstract}

%% file: chapters/introduction.tex
\section{\label{sec:intro}Introduction}
\begin{figure}[htb]
    \centering
    \includegraphics[width=0.49\textwidth]{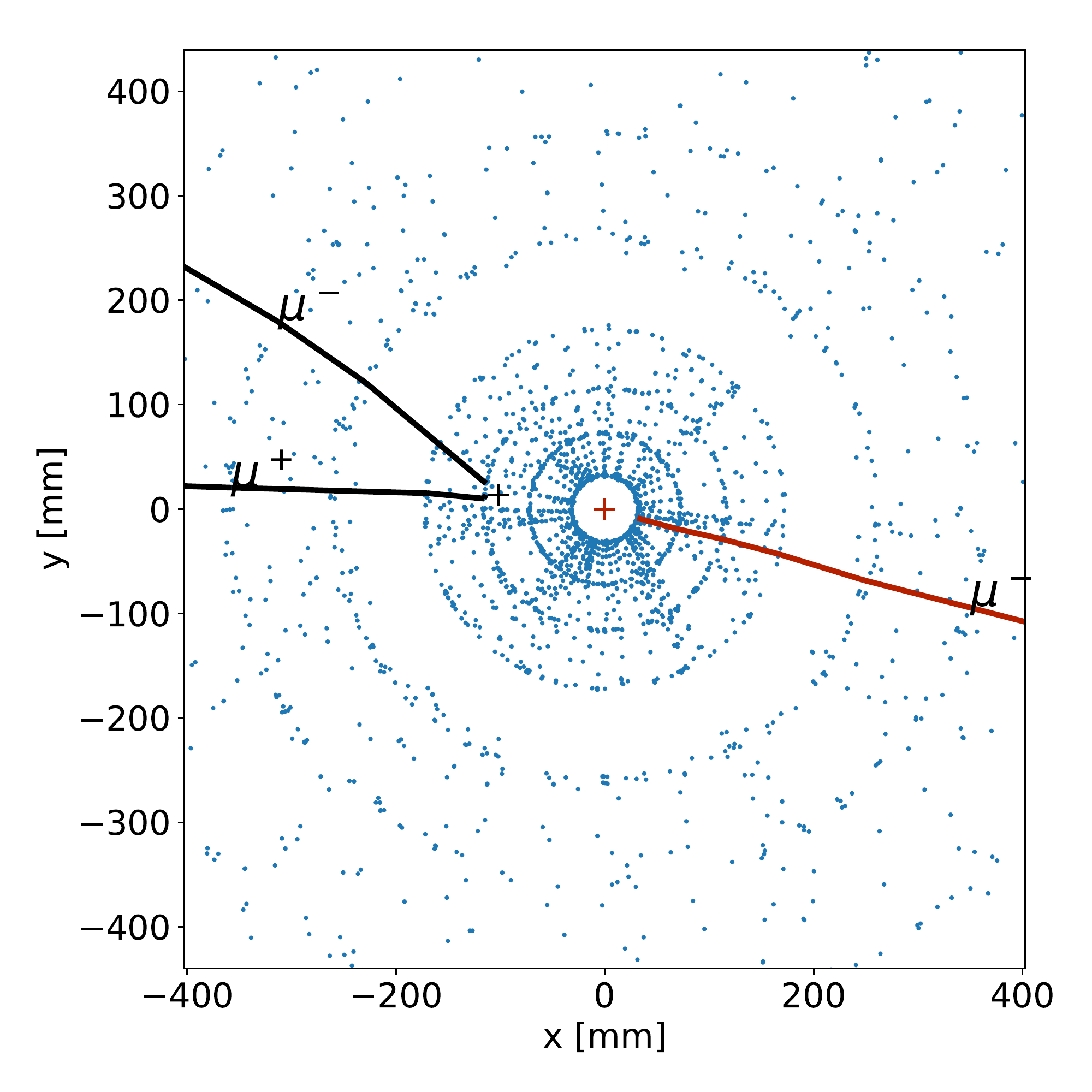}\hspace{2pc}
    \begin{minipage}[b]{14pc}
    \caption{Event display for a Heavy Neural Lepton event.
    Two muons in black result in two large radius tracks. One muon in red results in a prompt tracks.
    Tracks from underlying events are not displayed. Blue dots are detector recorded spacepoints.
    }
    \label{fig:event_display}
    \end{minipage}
\end{figure}

Various theories beyond the Standard Model (BSM) predict the existence of new particles with relatively long lifetimes that could be created in proton-proton collisions at the Large Hadron Collider~\cite{seesawHNL,Farrar:1978xj,PhysRevD.64.035002}. Such long-lived BSM particles can travel more than $\O(1~\text{mm})$ distance before decaying to SM particles, yielding unconventional signatures. For example, a right-handed Majorana neutrino (denoted by heavy neutral lepton HNL, or simply $N$) is proposed to address the fundamental question of the origins of neutrino masses~\cite{seesawHNL}. Depending on its mixing angles with neutrinos and mass parameters, the HNL may decay promptly or be long-lived. To facilitate our studies, we set the $N$ mass to be 15~GeV and its lifetime $c\tau = 100$~mm. Figure~\ref{fig:event_display} shows an event display of the HNL signature in a generic tracker detector. Each event is featured with one prompt track coming from the collision point and two displaced tracks coming from $N$. Prompt tracks point back to the interaction point (IP), while displaced tracks point to vertices that are away from IP, i.e. in a large radius in the transverse plane.

Standard tracking reconstruction algorithms are often optimized for prompt tracks. Stringent requirements are placed on the impact parameters of the reconstructed tracks  relative to the collision point to keep the computing time reasonable while still being very efficient in reconstructing prompt tracks with high purity~\cite{ATLAS:2017zsd, CMS:2014pgm}. As a result, standard tracking algorithms are inefficient in finding large radius tracks, which have large impact parameters. Therefore, experiments have to tune a dedicated configuration with relaxed requirements for efficiently finding large radius tracks, which inherently acquires additional computational cost. For example, Ref~\cite{ATLAS:2017zsd} observed the time needed to run the complete ATLAS event reconstruction including large radius tracking increases by about a factor of 2.5 with respect to the standard configuration.


Tested on the TrackML challenge dataset~\cite{trackml-challenge}, the Exa.TrkX pipeline showed promising computing and physics performance in reconstructing prompt tracks in High Luminosity-LHC-like collision events~\cite{heptrkx-ctd2017,Ju:2020xty,choma2020track,Ju:2021ayy}. This paper is to study the capability of the Exa.TrkX pipeline in finding large radius tracks. Section~\ref{sec:pipeline} briefly describes the Exa.TrkX pipleline, followed by Section~\ref{sec:data} detailing the data generation. Section~\ref{sec:results} shows the results on the trained pipeline, followed by the conclusion in Section~\ref{sec:conclusion}.




%% file: chapters/pipeline.tex
\section{E\lowercase{xa}.T\lowercase{rk}X Pipeline}
\label{sec:pipeline}

The Exa.TrkX pipeline is a Graph Neural Netwokr-based machine learning pipeline that reconstructs track candidates from a set of spacepoints~\footnote{spacepoints are 3D position measurements from a tracking detector.}. Details of the pipeline can be found in Ref~\cite{Ju:2021ayy}. 
The pipeline consists with three discrete stages, namely graph construction, edge classification, and track labeling.
Each stage exploits the relational information between two spacepoints based on their global positions. 

In the first stage, each event is constructed as a graph in that nodes are spacepoints and edges are connections between two spacepoints.
The graph construction method applies nearest neighbour search in a learned metric space, in which spacepoints from the same true track are close to each other and those from different tracks are far away, and filters out wrong connections through a multilayer perceptron (MLP). The metric learning~\cite{choma2020track} is implemented as a MLP, projecting the 3D spacepoint position to an 8D metric space. In this way, the metric learning focuses on the relational information between two spacepoints and does not know the tracking information. Therefore, it is insensitive to where tracks are created. 

Edge classification uses the Graph Network to assign a score ranging from 0 to 1 to each edge in the constructed graph. The higher the score is, the more likely the two spacepoints belong to the same track. There are many variations of Graph Network implemented for the Exa.TrkX pipeline~\footnote{https://github.com/HSF-reco-and-software-triggers/Tracking-ML-Exa.TrkX}. In this work, the Attention-based Graph Network originally developed in Ref~\cite{Farrar:1978xj} is used.

Track labelling stage reconstructs track candidates from the graph with edge scores. It uses a density-based clustering algorithm, DBSCAN~\cite{DBSCAN}, to cluster spacepoints into groups, and each group is considered as a track candidate. DBSCAN requires two parameters, epsilon and minPoints. Epslilon is the radius of the circle to be created around each data point to check the density and minPoints is the minimum number of data points required inside that circle. minPoints are set to 2 and epsilon is tuned by trying values from 0.1 to 1 with a step of 0.1. Epsilon of 0.4  gives highest track efficiency and track purity.

%% file: chapters/data_generation.tex
\section{Data Generation}
\label{sec:data}




The hard process is the heavy neutral lepton (HNL) production. 
\begin{equation} 
pp \rightarrow W \rightarrow \mu N, N \rightarrow \nu_\mu \mu^+ \mu^-,
\end{equation}
where $W$ is the Standard Model $W$ boson, $N$ represents heavy neutral particle with a mass of 15~GeV and a lifetime of 100 mm. 
The physics events are generated by Pythia~8~\cite{Sjostrand:2014zea} and simulated with the Fast Track Simulation~\cite{Edmonds:1091969} for a generic detector geometry implemented in the ACTS framework~\cite{Ai:2021ghi}. No additional interactions are included, i.e., pileup events are not included. We leave the study with different pileup conditions to the future.

This process creates three muons. A prompt muon coming from the $W$ boson leaves a prompt track. A pair of muons from the $N$ particle leaves two large radius tracks. On average, there are 140 tracks per event, most of which come from underlying interactions. All tracks are used to train the pipeline. Only prompt and displaced muons from the HNL process are used to calculate the tracking efficiency. We generated 10000 events for training and 5000 events for evaluation. No improvement was observed when more training events are used.


%

%% file: chapters/result.tex
\section{Result}
\label{sec:results}
\begin{figure}[htb]
    \centering
    \includegraphics[width=0.49\textwidth]{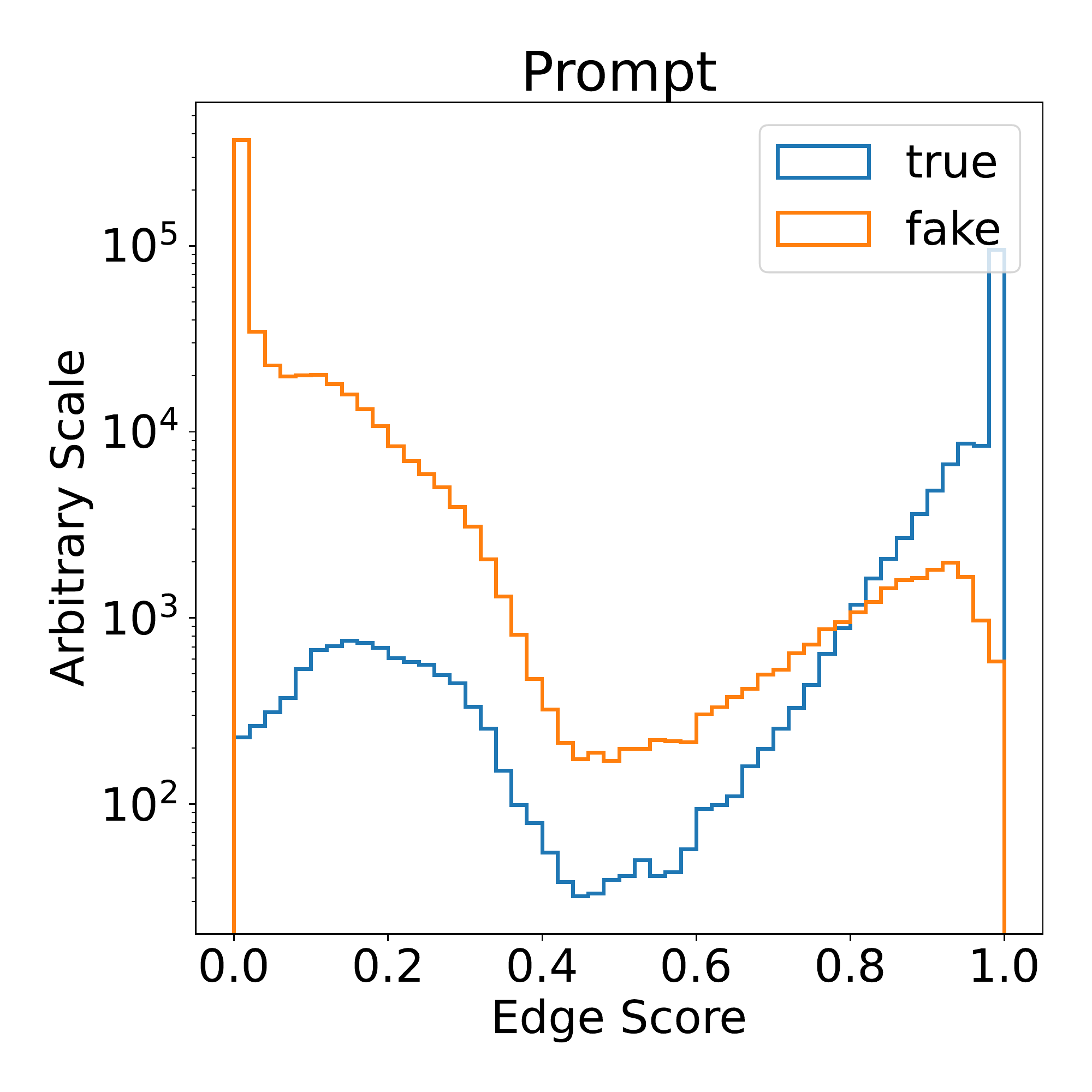}
    \includegraphics[width=0.49\textwidth]{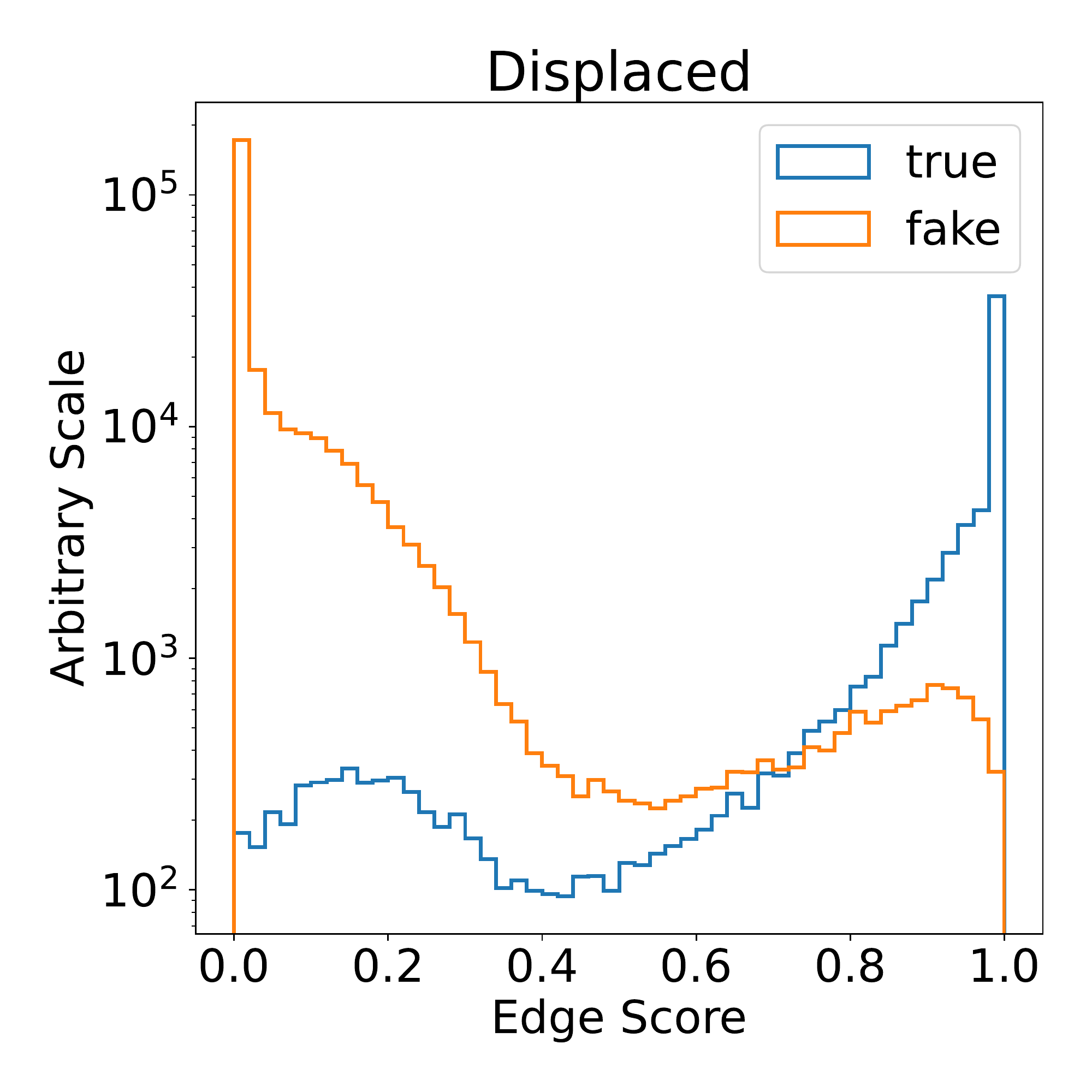}
    \caption{Edge score distribution from the Graph Neural Network for prompt tracks (Left) and displaced tracks (right).}
    \label{fig:gnn_scores}
\end{figure}

Graph construction is a crucial stage because it determines the overall performance of the pipeline. In an ideal graph, edges only connect two consecutive spacepoints coming from the same track, namely true edges. However, due to the imperfection of the graph construction algorithm, some edges connect two spacepoints coming from different tracks, namely fake edges. When training the pipeline, we monitor the edge efficiency and edge purity so as to determine how close the reconstructed graph is to the ideal graph. Edge efficiency is defined as the fraction of edges in ideal graphs that are kept in reconstructed graphs. Edge purity is defined as the fraction of edges in reconstructed graphs that connect two spacepoints coming from the same true track. The graph construction stage keeps 99.9 \% true edges with a purity of 5.7 \%. 
Figure~\ref{fig:gnn_scores} shows the distribution of edge scores of prompt and displaced tracks obtained from graph neural network, respectively.
With a thread of 0.5 on edge scores, the edge efficiency is 93.8~\% and edge purity is 87.0~\% for prompt tracks, and 92.5~\% and 84.7~\% for displaced tracks.

\begin{figure}[htb]
    \centering
    \includegraphics[width=0.49\textwidth]{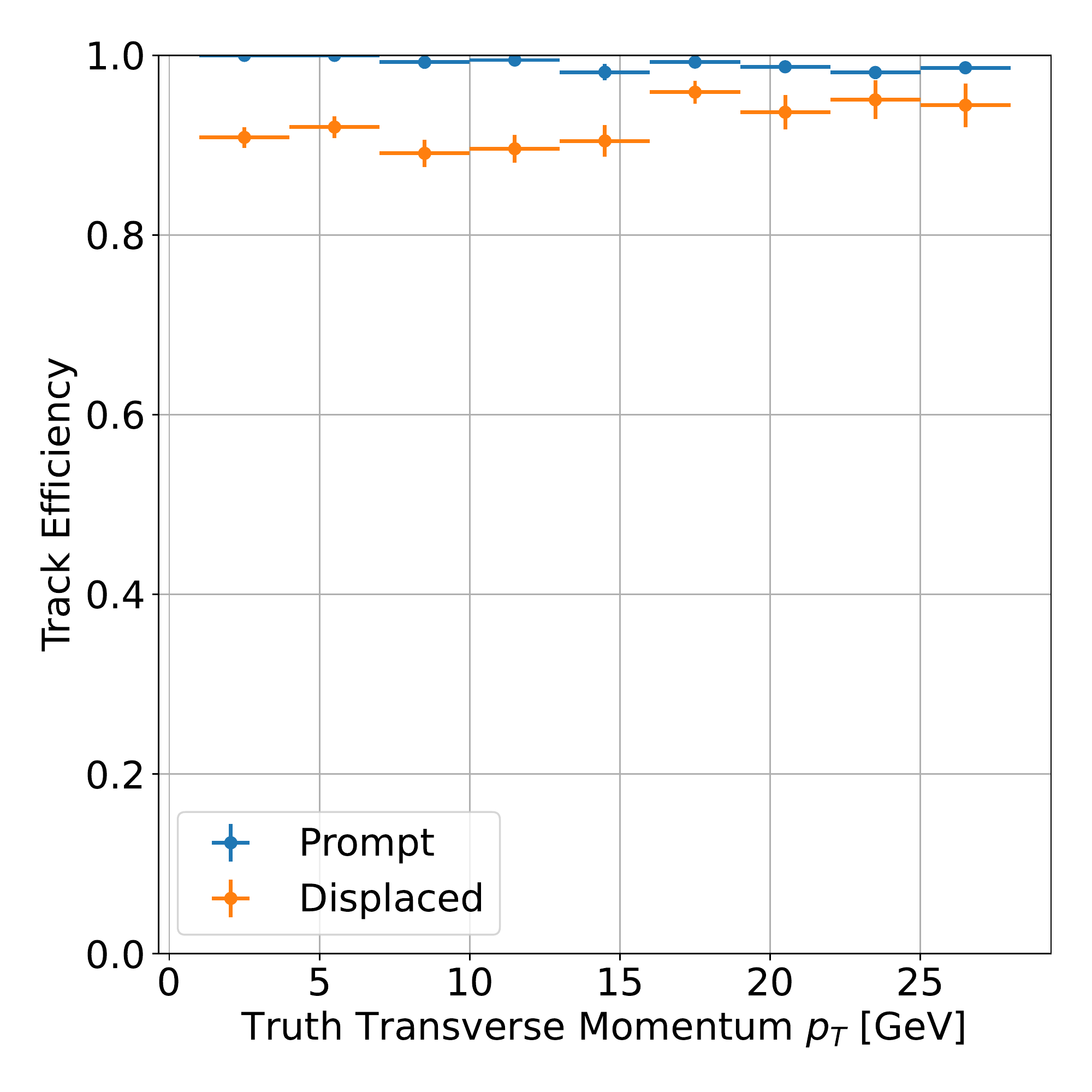}
    \includegraphics[width=0.49\textwidth]{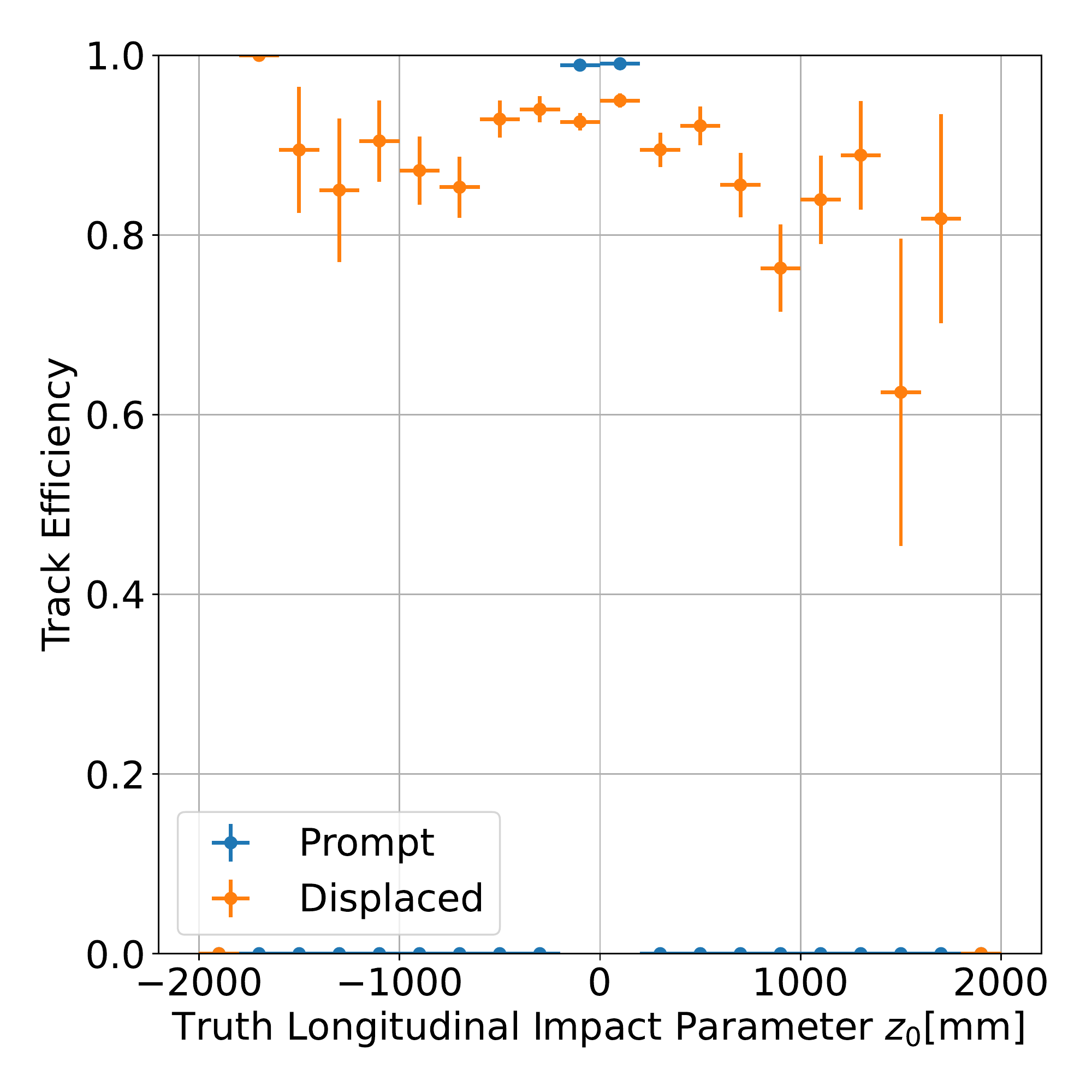}
    \includegraphics[width=0.49\textwidth]{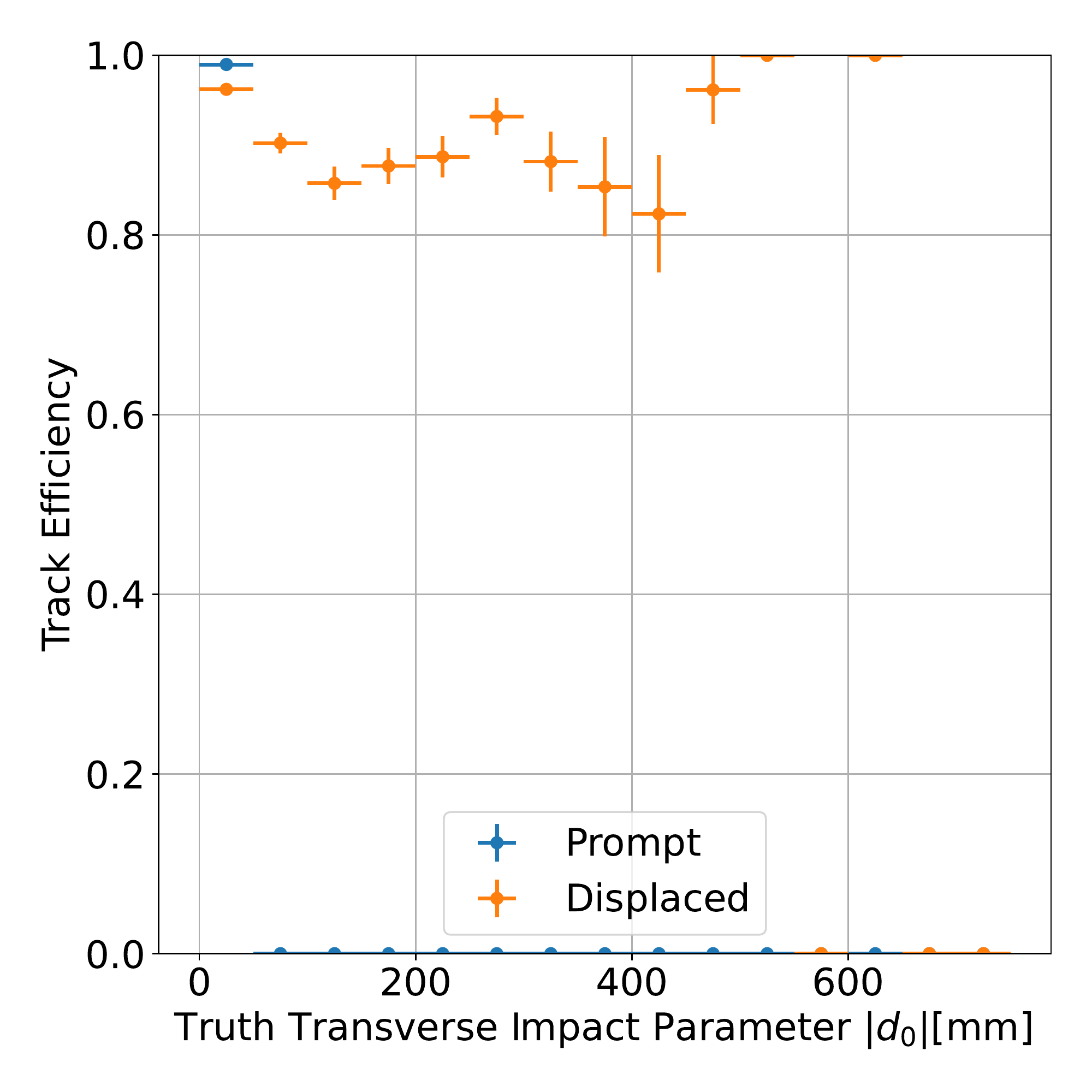}
    \includegraphics[width=0.49\textwidth]{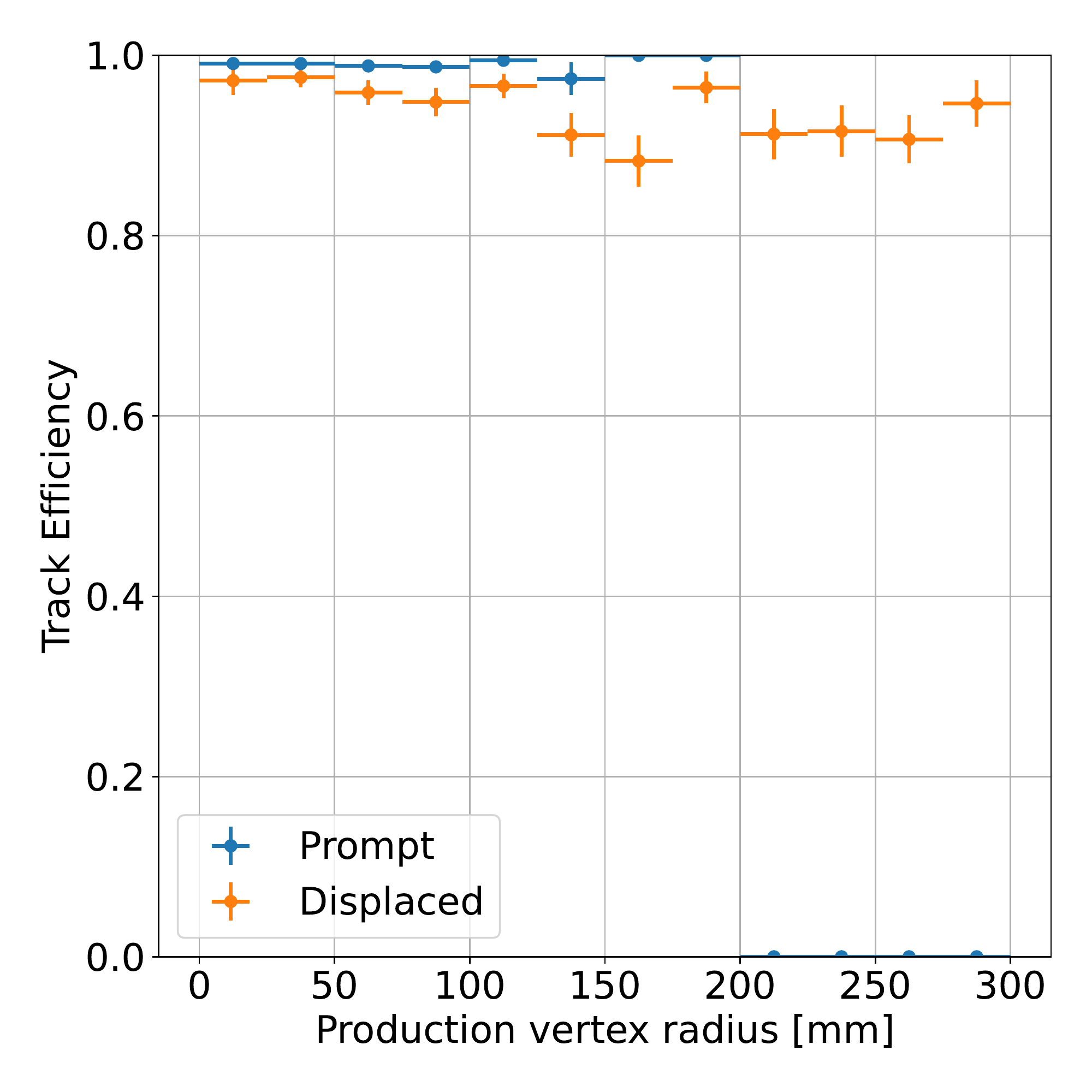}
    \caption{Track efficiency as a function of the true track $p_\text{T}$, impact parameters, and production vertex radius for prompt  tracks and displaced tracks.}
    \label{fig:tracking_eff}
\end{figure}

A true track is considered as reconstructable when the track $p_\text{T} > 1.0$~GeV and track $|\eta| < 4.0$ and it leaves at least five spacepoints in the detector. The tracking efficiency $\epsilon$ is defined as 
\begin{equation} 
\epsilon = \frac{N_{tracks}(\text{reconstructable, matched})}{N_{tracks}(\text{reconstrutable})}.
\end{equation}
A reconstructed track is matched to a true track if the number of spacepoints shared by the two tracks is greater than 50\% of total spacepoints in the reconstructed and true track, respectively and simultaneously. The track efficiency for all tracks in the event is $0.9561 \pm 0.0006$, and the efficiency for prompt tracks from the HNL process is $0.990 \pm 0.002$, and the efficiency for displaced tracks from the HNL process is $0.916 \pm 0.005$. Figure~\ref{fig:tracking_eff} shows the track efficiency as functions of truth transverse momentum $p_T$, truth transverse impact parameter $|d_0|$, truth longitudinal impact parameter $z_0$, and true production vertex radius.

%% file: chapters/conclusion.tex
\section{Conclusion}
\label{sec:conclusion}

Large radius tracks are unconventional signatures predicted by many Beyond the Standard Models. It is important to efficiently reconstruct those unconventional signatures without having significant additional computing time cost. We trained the Exa.TrkX pipeline to find prompt and large radius tracks in the proton-proton collision events. The trained pipeline reconstructs simultaneously both types of tracks with high efficiencies. Comparing with conventional track finding algorithms, the Exa.TrkX pipeline requires no additional computing cost in constructing large radius tracks. In addition, tracking algorithms using GNNs can take advantage of the capabilities of commercial accelerator devices such as GPUs, TPUs and FPGAs~\cite{Heintz:2020soy,Elabd:2021lgo}.  The pipeline may enable us to search for the new physics in real time. To that end, we will test the robustness of the pipeline in the presences of pileup events and for different physics processes. Furthermore, we will compare our results with conventional algorithms in terms of computing and physics performance.